\documentclass[11pt]{article}

\usepackage{amstext,amsgen,latexsym,amsmath}
\usepackage{amstext,amssymb,amsfonts,latexsym}
\usepackage{theorem} \usepackage{pifont}

 \newcommand{\hs}[1]{\hspace*{ #1 mm}}

 \def\bbox{\vrule height6pt width6pt depth1pt}

\theoremstyle{plain}
\theoremheaderfont{\bfseries}
 \newtheorem{theorem}{Theorem}[section] 
 \newtheorem{lemma}[theorem]{Lemma}
 
 \newtheorem{corollary}[theorem]{Corollary}
 \newtheorem{definition}[theorem]{Definition}

 \newenvironment{proof}{\par \noindent
            {\bf Proof. \hs{2}}}{\hfill$\Box$ \vspace*{3mm}}

 \newcommand{\ket}[1]{| #1 \rangle}

\newcommand{\ignore}[1]{}

\begin{document}

\title{Unbounded Error Quantum Query Complexity}
\author{{\sc Ashley Montanaro}$^1$, {\sc Harumichi Nishimura}$^2$ and {\sc Rudy Raymond}$^3$}
\maketitle
\begin{center}
$^1${Department of Computer Science, University of Bristol, UK} 

{\tt montanar@cs.bris.ac.uk}

$^2${School of Science, Osaka Prefecture University}

{\tt hnishimura@mi.s.osakafu-u.ac.jp}

$^3${Tokyo Research Laboratory, IBM Research, Japan}

{\tt raymond@jp.ibm.com}.
\end{center}
\begin{abstract}
This work studies the quantum query complexity of Boolean functions in a scenario 
where it is only required that the query algorithm succeeds with a probability 
strictly greater than $1/2$. We show that, just as in the communication complexity model, 
the unbounded error quantum query complexity is exactly half of its classical counterpart 
for any (partial or total) Boolean function. Moreover, we show that the ``black-box'' approach 
to convert quantum query algorithms into communication protocols by Buhrman-Cleve-Wigderson [STOC'98] 
is optimal even in the unbounded error setting.  

We also study a setting related to the unbounded error model, called the weakly unbounded error 
setting, where the cost of a query algorithm is given by $q+\log(1/2(p-1/2))$, 
where $q$ is the number of queries made and $p>1/2$ is the success probability of the algorithm. 
In contrast to the case of communication complexity, we show a tight $\Theta(\log n)$ separation
between quantum and classical query complexity in the weakly unbounded error setting for a
{\em partial} Boolean function. We also show the asymptotic equivalence between them for some
well-studied {\em total} Boolean functions. 
\end{abstract}

\section{Introduction}
Many models in computational complexity have several settings 
where different restrictions are placed on the success probability to evaluate a Boolean function $f$.
The most basic one is the {\em exact} setting: it requires that the computation 
of $f$ is always correct. In the polynomial-time complexity model, this corresponds to
the complexity class P. 
If we require the success probability to be only ``high'' (say, $2/3$), 
such a setting is called {\em bounded error}. The corresponding polynomial-time 
complexity class is known as BPP. The {\em unbounded error} setting is also standard. 
In this setting, it suffices to have a ``positive hint'', even infinitesimal, towards the right answer. 
That is, the unbounded error setting requires that the success probability 
to compute Boolean functions is strictly larger than $1/2$. The most famous 
model for this setting is also polynomial-time complexity, and PP is 
the corresponding complexity class. This setting also has connections with important concepts 
such as {\it polynomial threshold functions} in computational learning theory. 

There are two other major computing models which have been introduced to develop the lower bound method 
in complexity theory. The first one is the {\em communication complexity (CC)} model. 
The CC model measures the amount of communication for several parties, 
which have distributed inputs, to compute Boolean functions. 
The second one is the {\em query complexity (QC)} model. 
The QC model measures the amount of queries required for a machine 
with no input to compute a Boolean function by querying the input 
given in a black-box. 
The CC and QC models are also studied in the {\em quantum} setting, 
and there are many results on the performance gaps between classical 
and quantum computations (say, see \cite{Wol01}).  

So far, the unbounded error setting has also been studied in the CC and QC models. 
In the classical CC model, a large literature has developed since its introduction 
by Paturi and Simon \cite{PS86}. In the quantum case, Iwama et al.~\cite{INRY07,INRYb07} 
showed that the quantum CC of {\em any} Boolean function is almost half of its classical CC. 
Furthermore, a variant of the unbounded error setting was studied. 
It is often called the {\em weakly unbounded error} setting where 
the cost can be defined by $q+\log(1/2(p-1/2))$, where $q$ is the number 
of communication (qu)bits and $p>1/2$ is the success probability\footnote{Some previous work does
not have the factor of 2 in the denominator; see Section \ref{definitions} for a discussion.}. 
Recently, Buhrman et al.~\cite{BVW07} and Sherstov \cite{She07} independently 
showed that there is a Boolean function that exponentially separates 
the classical weakly unbounded error CC and unbounded error CC, which solves an open problem that
remained from \cite{BFS86}. 
On the other hand, there can only be a constant gap between quantum and classical CCs 
for Boolean functions in the weakly unbounded error setting \cite{INRYb07,Kla07}. 

The study of the QC model in the unbounded error setting has been developed 
implicitly as the study of the sign-representing polynomial (say, \cite{ABFR94,Bei94}) 
since Beals et al.~\cite{BBCMW98} gave the nice characterization of the (quantum) QC 
by polynomials. In fact, Buhrman et al.~\cite{BVW07} mentioned the close relation 
between sign-representing polynomials and QCs of Boolean functions.  
However, there is no explicit literature on unbounded error quantum QCs, such as 
the relationship to classical QC and the weakly unbounded error variants. 

{\bf Our Results.~}
In this paper we deal with the unbounded error quantum QC and study 
its relations to the other unbounded error concepts. 
First, we show that, as in the case of CC, the unbounded error quantum QC 
of some (total/partial) Boolean function is always exactly half 
of its classical counterpart. Second, we discuss the relation between the unbounded error quantum QC and CC. There is a famous result by Buhrman et al.~\cite{BCW98} that ``reduces'' quantum CC to quantum QC, 
which is a ``black-box'' approach to convert quantum query algorithms into communication protocols 
with $O(\log n)$ overhead. Combining with results on unbounded error quantum CC, we show that the overhead of the black-box approach of \cite{BCW98} is optimal, that is, $\Omega(\log n)$ overhead is inevitable. Moreover, we see that this bound on overhead factor holds under several other settings. Third, we show an unbounded separation, $O(1)$ vs.~$\Omega(\log n)$, between the weakly unbounded error quantum and classical QCs for some {\em partial} function. We can easily see that the gap is tight. On the other hand, we show that the separation is only constant for some well-known {\em total} Boolean functions such as PARITY, AND, OR and threshold functions. Finally, we show that weakly unbounded error QC can be exponentially smaller than a well-studied complexity measure of Boolean functions, the average sensitivity.

{\bf Related Work.~} In a similar direction, de Wolf \cite{Wol03} 
characterized the nondeterministic\footnote{This could be called 
{\it one-sided unbounded error} since the computation is required to be exact in one side of the output.} 
quantum QC and CC by, respectively, {\it nondeterministic degree} of approximating polynomials 
and {\it nondeterministic rank} of communication matrices. When comparing classical and quantum complexities 
under these models, de Wolf showed strong separations; an unbounded gap for QC and an exponential gap for CC 
(while the first unbounded gap for CC was shown before in \cite{Mass01}).  
Under a different (i.e., certificate based) type of nondeterminism, 
a quadratic separation between quantum and classical CC is known for some total function \cite{Gall06}. 

{\bf Organization of the paper.~} We begin, in Section \ref{definitions}, by giving the formal definitions of the models that we discuss in this paper. Section \ref{sec:uq-uc} contains the relation between 
unbounded error quantum and classical QCs. 
In Section \ref{treduction}, we show the optimality of the reduction of \cite{BCW98} 
from quantum CC to quantum QC. In Section \ref{wubq}, 
we compare the weakly unbounded error quantum QC to other several QCs 
and to average sensitivity. The paper finishes with some concluding remarks.

\section{Definition and Models}\label{definitions}
We first list some useful definitions, starting with unbounded error polynomials.

\begin{definition}\label{udeg} 
Let $f:\{0,1\}^n \rightarrow \{0,1\}$ be a Boolean function of $n$ variables, 
and $q:\{0,1\}^n \rightarrow [0,1]$ be a real multilinear polynomial. We say that 
$q$ is an {\em unbounded error polynomial for $f$} if for any $x\in\{0,1\}^n$, 
$q(x)>1/2$ if $f(x)=1$ and $q(x)<1/2$ if $f(x)=0$. 
We denote the lowest degree among all unbounded error polynomials for $f$ as ${udeg}(f)$.
\end{definition}

Note that this definition is given in terms of total Boolean functions, 
but we can naturally extend it to partial functions. 
Throughout this paper, we use the term ``Boolean functions'' 
for results that hold for both partial and total functions; if not, we mention it explicitly.

There have been many studies and extensive results in the literature on polynomials 
that {\it sign-represent} Boolean functions \cite{ABFR94,Bei94,MP88}. 
A polynomial $p:\{0,1\}^n \rightarrow \mathcal{R}$ is said to sign-represent $f$ 
if $p(x) > 0$ whenever $f(x) = 1$, and $p(x) < 0$ whenever $f(x) = 0$~\footnote{Note that in the literature, 0/1 is usually replaced by 1/-1 for convenience.}. 
If $|p(x)| \le 1$ for all $x$, we say that $p$ is {\it normalized}. 
The {\em bias} of a normalized polynomial $p$ is defined as $\beta = \min_x |p(x)|$. 
Denoting the minimum degree of polynomials that sign-represent $f$ as ${sdeg}(f)$, 
it is easy to see that ${udeg}(f) = {sdeg}(f)$, since an unbounded error polynomial $q$ for $f$ 
can be obtained from a sign-representing polynomial $p$ for $f$ as follows: 
$q(x) = 1/2 + p(x)/2$. In the following, we will use some results about polynomials that sign-represent $f$ 
to characterize the unbounded error QC. 

It is folklore that every real multilinear polynomial $q$ of degree at most $d$ can be represented 
in the so-called {\it Fourier basis}. Namely, 
\begin{equation}\label{fourierbasis}
q(x) = \sum_{S\in S_d} \hat{q}(S) (-1)^{x_S}, 
\end{equation}
where $S_d$ denotes the set of all index sets $S\subseteq[n]$ of at most $d$ variables, 
and $x_S$ denotes the XOR (or parity) of the bits of $x$ on index set $S$, namely, $x_S = \oplus_{i \in S} x_i$. 

Next, we give the definitions of unbounded error QC and CC, as well as their weak counterparts. 
In Section \ref{treduction}, we will also mention a result that is related to the so-called 
{\it nondeterministic} QC and CC whose details can be found in \cite{Wol03}. 

\begin{definition}\label{uq} Let ${UQ}(f)$ and ${UC}(f)$ be 
the unbounded error quantum and classical, respectively, QCs of a Boolean function $f$. 
Namely, ${UQ}(f)$ (resp.~${UC}(f)$) is the minimum number of quantum (resp.~classical) queries to a black box 
that holds the input $x$ to $f$ such that $f$ can be computed with a success probability greater than $1/2$. 
Let ${UQC}(g)$ and ${UCC}(g)$ be the unbounded error quantum and classical, respectively, CCs 
of a Boolean function $g:\{0,1\}^{n_1} \times \{0,1\}^{n_2} \rightarrow \{0,1\}$.

Define the {\em bias} $\beta$ of a quantum or classical query algorithm (resp.~communication protocol) 
which succeeds with probability $p>1/2$ as $p-1/2$. Then the {\em weakly unbounded cost} of such an algorithm 
(resp.~protocol) is equal to the number of queries (resp.~communicated bits or qubits) 
plus $\log 1/2\beta$\footnote{In this paper, $\log$ denotes the logarithm taken to base 2.}. 
Let ${WUQ}(f)$, ${WUC}(f)$, ${WUQC}(g)$ and ${WUCC}(g)$ 
be the {\it weakly} unbounded error counterparts of the previous measures, 
given by the minimum weakly unbounded cost over all quantum or classical query algorithms 
and communication protocols, respectively.
\end{definition}

Note that in the above definition ${UQC}$ and ${UCC}$ refer to two-way CC. 
However, since two-way CC only differs from one-way CC by at most one qubit or bit \cite{PS86,INRYb07}, 
for simplicity we will mainly use results in one-way CC, which have been much studied in \cite{PS86,INRY07}.
Also note that some previous work defines the weakly unbounded cost as the number of queries plus
$\log 1/\beta$ \cite{BVW07,INRYb07}. However, we prefer the present definition, as it
ensures that weakly unbounded query/communication complexity is never greater than exact
query/communication complexity. For example, with the previous definition, a function $f$ for which
there exists an optimal classical algorithm that uses one query and succeeds with certainty would have
${WUC}(f)=2$, whereas with the present definition ${WUC}(f)=1$.

\section{Unbounded Error Quantum and Classical QCs}\label{sec:uq-uc}

In \cite{INRY07}, it is shown that ${UQC}(f)$ is always exactly half of ${UCC}(f)$ 
for any (partial or total) Boolean function $f$, and 
that both complexity measures can be characterized in terms of {\em arrangements} (see \cite{INRY07} for definitions relating to arrangements).

\begin{theorem}\label{ucc} Let $k_f$ denote 
the smallest dimension of an arrangement that realizes $f$. 
Then ${UQC}(f) = \lceil \log{(k_f+1)}/2 \rceil$, and ${UCC}(f) = \lceil \log{(k_f+1)} \rceil$.
\end{theorem}

In the following, we will show that in the unbounded error QC model, 
the equivalent result -- that quantum query complexity is always exactly half of its classical counterpart -- 
also holds for any Boolean function. For this purpose, we need the following lemmas.

The first lemma gives a lower bound on the number of queries 
in terms of the minimum degree of representing polynomials. 

\begin{lemma}\label{degquery}[Beals et al. \cite{BBCMW98}] 
The amplitude of the final basis states of a quantum algorithm using $T$ queries can be 
written as a multilinear polynomial of degree at most $T$. 
\end{lemma}

The second lemma gives an exact quantum algorithm for computing the parity of $n$ variables with just $n/2$ queries. 

\begin{lemma}\label{parity}[Farhi et al. \cite{FGGS98}] Let $S \subseteq [n]$ be a set of indices of variables. 
There exists a quantum algorithm for computing $x_S$ with $\lceil |S|/2 \rceil$ queries. 
That is, there exists a unitary transformation $U_f$ which needs exactly $\lceil |S|/2 \rceil$ queries: 
for any $b \in \{0,1\}$, 
$$
U_f\ket{S}\ket{0^m}\ket{b} = \ket{S}\ket{\psi_S}\ket{b \oplus x_S},
$$
where $\ket{0^m}$ and $\ket{\psi_S}$ are the workspace quantum registers before 
and after the unitary transformation, respectively.
\end{lemma}

The third lemma was shown recently by Buhrman et al. \cite{BVW07} and turns out to be very useful 
in characterizing the unbounded error QC of Boolean functions. 

\begin{lemma}\label{probcoeff}[Buhrman et al. \cite{BVW07}]
Suppose that there exists a multilinear polynomial $p$ of $d$-degree that sign-represents 
$f:\{0,1\}^n \rightarrow \{0,1\}$ with bias $\beta$. Define $N=\sum_{i=0}^d \binom{n}{i}$. 
Then there also exists a multilinear polynomial $q(x) = \sum_{S \in S_d} \hat{q}(S) (-1)^{x_S}$ 
of the same degree and bias $\beta/\sqrt{N}$ that sign-represents $f$ such that 
$\sum_{S \in S_d} |\hat{q}(S)| = 1$. 
\end{lemma}

Now we are ready to state the relationship between $UQ(f)$ and $UC(f)$ for Boolean functions $f$. 
\begin{theorem}\label{uqhalf}
For any Boolean function $f:X \rightarrow \{0,1\}$ such that $X \subseteq \{0,1\}^n$, it holds that: 
$$
{UQ}(f) = \left\lceil \frac{{UC}(f)}{2} \right\rceil = \left\lceil \frac{{udeg}(f)}{2} \right\rceil.
$$
\end{theorem}

\begin{proof}
[${UC}(f) = {udeg}(f)$]~It follows from a result in Buhrman et al. \cite{BVW07} 
that an unbounded error randomized algorithm for $f$ using $d$ queries 
is equivalent to a $d$-degree polynomial $p$ that sign-represents $f$, 
and hence to a $d$-degree unbounded error polynomial $q$ for $f$. 

[${UQ}(f) \ge {udeg}(f)/2$]~Let ${\cal A}$ be an unbounded error quantum algorithm for $f$ 
using $UQ(f)$ queries. Note that the acceptance probabilities of quantum algorithms can be written 
as the sum of the absolute values squared of the amplitude magnitudes of the corresponding basis states. 
By Lemma~\ref{degquery}, the acceptance probability of ${\cal A}$ can be written 
as a multilinear polynomial of degree at most $2UQ(f)$. Hence, ${udeg}(f) \le 2{UQ}(f)$. 

[${UQ}(f) \le \lceil {udeg}(f)/2 \rceil$]~This follows from Lemmas~\ref{probcoeff} and \ref{parity}. 
First, let $\delta(y) = 1$ if $y > 0$, and $\delta(y) = 0$ otherwise. 
With regard to the Fourier representation of polynomial $p$ that sign-represents $f$ 
as in Eq.~(\ref{fourierbasis}), and for a fixed $x \in \{0,1\}^n$, we can write
$$
p(x) = \sum_{S\in S_{udeg(f)}} \hat{p}(S) (-1)^{x_S} 
= \sum_{S\in S_x^+}|\hat{p}(S)|-\sum_{S\in S_x^-}|\hat{p}(S)|
$$
such that $S_{x}^{+} = \{S| x_S \oplus \delta(\hat{p}(S)) = 1\}$, 
and $S_{x}^{-} = \{S | x_S \oplus \delta(\hat{p}(S)) = 0\}$. 
By Lemma~\ref{probcoeff}, we can assume that $\sum_{S\in S_{udeg(f)}}|\hat{p}(S)|=1$. 
Then, we have $\sum_{S\in S_{x}^{+}} |\hat{p}(S)| > 1/2$ if $f(x) = 1$, 
and $\sum_{S\in S_{x}^{+}} |\hat{p}(S)| < 1/2$ otherwise. 
Thus, the unbounded error quantum algorithm for $f$ can be obtained 
by computing the XOR of $x_S$ and $\delta(\hat{p}(S))$: the former 
by applying Lemma~\ref{parity} with $\lceil |S|/2 \rceil\leq \lceil udeg(f)/2\rceil$ 
queries, and the latter without query cost, as summarized in the following steps. 
\begin{enumerate}
\item Prepare quantum state $\ket{\psi_p} = 
\sum_{S\in S_{udeg(f)}} \sqrt{|\hat{p}(S)} \ket{S}\ket{0^m}\ket{\delta({\hat{p}(S))}}$. 
\item Apply the unitary transformation $U_f$ of Lemma~\ref{parity} for obtaining the parity of $x$ 
on index set $S$, whose result is stored in the last register. The quantum state after the transformation is: 
$$
U_f\ket{\psi_p} = 
\sum_{S\in S_{udeg(f)}} \sqrt{|\hat{p}(S)} \ket{S}\ket{\psi_S}\ket{\delta({\hat{p}(S))}\oplus x_S}.
$$
\item Measure the last register, and output the result of the measurement. 
\end{enumerate}

This completes the proof. 
\end{proof}

Thus, we can conclude that the unbounded error quantum QC is always exactly half of its classical counterpart. 
The following corollary is immediately obtained from Theorem \ref{uqhalf} and Corollary 9 of \cite{DS03}.

\begin{corollary} Almost every function $f:\{0,1\}^n\rightarrow\{0,1\}$ 
has unbounded error quantum QC $n/4 \le {UQ}(f) \le n/4 + O(\sqrt{n\log{n}})$. 
\end{corollary}

\section{Tightness of Reducing CC to QC}\label{treduction}

Buhrman, Cleve and Wigderson~\cite{BCW98} gave a method for reducing a quantum communication protocol 
to a quantum query algorithm with $O(\log{n})$ overhead, which we call the {\it BCW reduction}. 
In brief, it is proven that if the QC of a quantum algorithm of $f$ on input $x$ with length $n$ is $T$, 
then the corresponding CC of $f$ on distributed input $a$ on Alice 
and $b$ on Bob such that $x = a \oplus b$ (or any other bitwise function) is at most $O(T \log{n})$. 
In the reverse direction, this implies that any lower bound $C$ in the CC side 
is translated into a lower bound $\Omega(C/\log{n})$ in the QC side. 
The BCW reduction is exact: the success probability of the communication protocol 
is the same as that of the query computation. 
In fact, \cite{BCW98} proved some interesting results using this reduction, such as the first non-trivial 
quantum protocol for the disjointness problem, which used $O(\sqrt{n}\log n)$ communication 
by a reduction from Grover's quantum search algorithm. 
Later, this upper bound was improved to $O(\sqrt{n} \log^{*}(n))$ by \cite{HW03}, 
and finally to $O(\sqrt{n})$, which matches the lower bound shown in \cite{Raz02}, by \cite{AA05} 
with ingenious simulation techniques. However, unlike the results of \cite{BCW98}, 
those techniques seem to be limited only to specific functions such as disjointness. 

Thus, it is of interest to know whether there exists a universal reduction 
similar to the BCW reduction but with $o(\log{n})$ overhead and preserving the success probability. 
This might be achieved by designing new reduction methods. 
In addition, smaller overhead might be achieved by relaxing the success probability condition, 
that is, allowing the success probability of the resulting protocol to be significantly lower than 
that of the original algorithm. To look for the possibility of such a universal reduction, 
one can consider relations between quantum QC and CC by using such a reduction as a ``black-box'' 
under various settings for the required success probability.

Our result in this section is the optimality of the BCW reduction 
in the exact, nondeterministic, and unbounded error settings. 
We show the existence of Boolean functions whose quantum QCs are constant, 
while the CCs of their distributed counterparts are $\Omega(\log n)$, where $n$ is the input length. 
We start with presenting a partial Boolean function, which is a variant of the Fourier Sampling problem 
of Bernstein and Vazirani \cite{BV97}, for showing the optimality of the BCW reduction under the exact 
and nondeterministic settings. 

\begin{definition} 
For $x,r \in \{0,1\}^m$, let $F^r$ be a bit string of length $n=2^m$ 
whose $x$-th bit is $F^r_x = \sum_i x_i \cdot r_i \mod 2$. 
Let also $g$ be another bit string of length $n$. 
The {\em Fourier Sampling} (FS) of $F^r$ and $g$ is defined by $\mathrm{FS}(F^r,g) = g_r$. 
When Alice and Bob are given $(F^a,g)$ and $(F^b,h)$, respectively, as their inputs 
where $a,b\in\{0,1\}^m$ and $g,h\in\{0,1\}^n$, the {\em Distributed Fourier Sampling} (DFS) 
on their inputs is $\mathrm{DFS}((F^a,g),(F^b,h)) = \mathrm{FS}(F^a\oplus F^b,g \oplus h)$.
\end{definition}

{\bf Remark.~} FS can also be considered as a variant of the Goldreich-Levin problem 
in the cryptographic setting for {\it noisy} $F^r$ and a bit string $g$ such that $g_i = 1$ 
if and only if $i = r$, see, e.g., the lecture note by Bellare \cite{Bell99}.

Now we can construct the following two-query quantum algorithm for $\mathrm{FS}$: 
Given input $(F^r,g)$ (in a black-box), 
(i) Determine $r$ with one query to $F^r$ with certainty
by the quantum algorithm of \cite{BV97}. (ii) Output $\mathrm{FS}(F^r,g)=g_r$ with
one query to $g$. Thus, we have the following lemma.

\begin{lemma}\label{eqfs}
The exact (and hence nondeterministic) quantum QC of $\mathrm{FS}$ is $2$.
\end{lemma}

Now, we can apply the BCW reduction to the quantum algorithm of Lemma 
\ref{eqfs} to obtain an exact (or nondeterministic) quantum protocol for computing $\mathrm{DFS}$ with $O(\log n)$ qubits communication cost. However, it can be shown that we cannot compute $\mathrm{DFS}$ better than the BCW reduction. 
For this purpose, we consider the set of inputs $((F^a,g),(F^b,h))\in (\{0,1\}^n)^2
\times(\{0,1\}^n)^2$ such that $g\oplus h=10^{n-1}$. 
For such inputs, $\mathrm{DFS}((F^a,g),(F^b,h))=1$ (resp.~$0$) implies $a=b$ (resp.~$a\neq b$)
since $\mathrm{DFS}((F^a,g),(F^b,h))=\mathrm{FS}(F^a\oplus F^b,g\oplus h)=\mathrm{FS}(F^{a\oplus b},10^{n-1})
=(10^{n-1})_{a\oplus b}$ (the $a\oplus b$-th bit of $10^{n-1}$). 
This means that if the nondeterministic CC of $\mathrm{DFS}$ is $o(\log n)$, 
then that of $\mathrm{EQ}_{\log n}$ (the equality on two $\log n$  bits $a$ and $b$)
is also $o(\log n)$, which contradicts the fact that the nondeterministic quantum CC 
of $\mathrm{EQ}_{\log n}$ is $\Omega(\log n)$ \cite{Wol03}. Thus, we obtain the following result.

\begin{lemma}\label{eqdfs} 
The nondeterministic (and hence exact) quantum CC of $\mathrm{DFS}$ is $\Omega(\log n)$. 
\end{lemma}

Next, we consider the reduction in the unbounded error setting, 
that is, the possibility of converting a quantum query algorithm 
into the corresponding communication protocol with $o(\log n)$ overhead, 
but the success probability of the resulting protocol becomes very close to half.  
It turns out that even in this setting, the BCW reduction is optimal. 
In fact, we show that the ODD-MAX-BIT function, 
a total Boolean function introduced in \cite{Bei94}, has QC $O(1)$ 
while its distributed version has CC $\Omega(\log{n})$. 
To this end, we give the definition of $\mbox{ODD-MAX-BIT}_n$ (or $\mbox{OMB}_n$ for short) 
and its distributed variant $\mbox{DOMB}_n$. 

\begin{definition}
For any $x \in \{0,1\}^n$, let us define $\mathrm{OMB}_n(x) = k \mod 2$, 
where $k$ is the largest index of $x$ such that $x_k = 1$ ($k = 0$ for $x = 0^n$). 
For any $a,b \in \{0,1\}^n$, let us also define $\mathrm{DOMB}_n(a,b) 
=\mathrm{OMB}_n(a\wedge b)$, where $a\wedge b$ is the bitwise AND of $a$ and $b$. 
\end{definition}

For the functions OMB and DOMB, we show the following theorem, which shows that 
the BCW reduction cannot be improved in the unbounded error setting. 
Note that \cite{BVW07} proved that ${UQC}(\mbox{OMB}_n) \le \lceil{\log{n}}\rceil + 1$. 

\begin{theorem}\label{omb_domb}
${UQ}(\mathrm{OMB}_n) = 1$ and ${UQC}(\mathrm{DOMB}_n) \ge \frac{\log{n}-3}{2}$. 
\end{theorem}

\begin{proof}
The proof for ${UQ}(\mbox{OMB}_n) = 1$ is easy. 
Let us consider the following classical algorithm: 
Query $x_i$ with probability $p_i = \frac{2^i}{2^{n+1}-2}$. 
Then, output $i \mod 2$ if $x_i = 1$, and the result of a random coin flip if $x_i=0$. 
It can be seen that the success probability is always bigger than $1/2$ for all positive integers $n$. 

The bound for ${UQC}(\mbox{DOMB}_n) \ge \frac{\log{n}-3}{2}$ follows from 
the lower bound on quantum random access coding (which is also 
known as the INDEX function) shown in \cite{INRY07,INRYb07}. 
For $a \in \{0,1\}^n$ and $b \in \{0,1\}^{\log{n}}$, $\mbox{INDEX}_n(a,b)$ 
is defined as the value of the $b$-th bit of $a$, or $a_b$. 
Then, we consider the case that Alice uses $x = a_10a_20\ldots a_n0$, and 
Bob uses $y = y_1y_2y_3\ldots y_{2n}$ such that $y_{j} = 1$ iff $j = 2b -1$, 
as inputs to the protocol for $\mbox{DOMB}_{2n}$. 
Clearly, $\mbox{DOMB}_{2n}(x,y) = \mbox{INDEX}_n(a,b)$. 
However, according to \cite{INRY07,INRYb07}, ${UQC}(\mbox{INDEX}_n) \ge \frac{1}{2}\log(n+1) - 1$. 
Therefore, $UQC(\mbox{DOMB}_n)\geq \frac{1}{2}\log(n/2+1)-1\geq (\log n-1)/2-1=(\log n-3)/2$. 
\end{proof}

Now we can summarize the results in this section by Theorem \ref{bcw_tight}: 
Here, for any Boolean function $F:\{0,1\}^n\rightarrow\{0,1\}$, 
the {\em distributed function} of $F$ induced by the bitwise XOR (resp.~AND), 
denoted by $F^{\oplus}$ (resp.~$F^{\wedge}$), is defined 
by $F^{\oplus}(x,y)=F(x\oplus y)$ (resp.~$F^{\wedge}(x,y)=F(x\wedge y)$). 

\begin{theorem}\label{bcw_tight}
The following hold:

(1) Assume that there is a procedure $\mathcal{A}$ that, for any function $F:\{0,1\}^n\rightarrow\{0,1\}$, 
converts a nondeterministic (exact, resp.) quantum algorithm for $F$ using $T(n)$ queries into 
a nondeterministic (exact, resp.) quantum communication protocol for $F^{\oplus}$ using $O(D(n)T(n))$ qubits. 
Then, $D(n) = \Omega(\log{n})$.  

(2) Assume that there is a procedure $\mathcal{A}$ that, for any function $F:\{0,1\}^n\rightarrow\{0,1\}$, 
converts an unbounded error quantum algorithm for $F$ using $T(n)$ queries into 
an unbounded error quantum communication protocol for $F^{\wedge}$ using $O(D(n)T(n))$ qubits. 
Then, $D(n) = \Omega(\log{n})$.  
\end{theorem}

As a direct consequence of Theorem \ref{bcw_tight}(2), we can derive a general and tight relation 
between ${udeg}(F)$ and the dimension of an arrangement $k_F$ 
that realizes the distributed version of $F$ (see Theorem \ref{ucc}).

\begin{corollary}
For any Boolean function $F:\{0,1\}^n \rightarrow \{0,1\}$, 
${udeg}(F) = \Omega(\log{k_{F^{\wedge}}}/\log{n})$. 
Moreover, the above relation is tight for some $F$ and $F^{\wedge}$.
\end{corollary}

By the result of \cite{Wol03} and Theorem \ref{bcw_tight}(1), 
the above corollary also holds in the nondeterministic setting 
by replacing $F^{\wedge}$, ${udeg}$ and $\log{k}$ by $F^{\oplus}$, 
${ndeg}$ (or the degree of nondeterministic polynomials) 
and $\log{{nrank}}$ (or the log of the rank of nondeterministic communication matrix), respectively. 

\section{Weakly Unbounded Error Quantum and Classical QCs}\label{wubq}

In this section, we study the weakly unbounded error QC. 
First, we observe an unbounded gap, $O(1)$ vs. $\Omega(n^{1/3}/\log n)$,  
between unbounded error and weakly unbounded error quantum QCs. 
Next, we consider gaps between quantum and classical weakly unbounded error QCs 
for partial and total functions. 

\subsection{Unbounded Gaps between ${UQ}$ and ${WUQ}$}
As mentioned in Section 1, an exponential gap between ${UQC}$ and ${WUQC}$ in the CC model 
was shown by Buhrman et al.~\cite{BVW07} and Sherstov \cite{She07}. In \cite{BVW07} 
the function $\mbox{DOMB}_n$ was used to show the gap. We can easily see that by using $\mbox{OMB}_n$ 
a similar (but unbounded) gap is shown also in the QC model.

\begin{lemma}
${UQ}(\mathrm{OMB}_n) = 1$ and ${WUQ}(\mathrm{OMB}_n) = \Omega(n^{1/3}/\log{n})$.
\end{lemma}

\begin{proof}
${UQ}(\mathrm{OMB}_n)=1$ was already shown in Theorem \ref{omb_domb}. 
The lower bound of ${WUQ}(\mbox{OMB}_n)$ follows from the result 
${WUQC}(\mbox{DOMB}_n) = \Omega(n^{1/3})$ in \cite{BVW07} 
combined with the BCW reduction.
\end{proof}

\subsection{Tight Gaps between $WUQ$ and $WUC$ for Partial Functions}\label{tgaps}

In Section \ref{sec:uq-uc}, we showed that there is always a gap between unbounded error quantum and 
classical QCs. Namely, the former is always exactly half as large as the latter, 
which is similar to the CC model \cite{INRY07,INRYb07}. In fact, in the CC model,
weakly unbounded error quantum CC follows the same pattern:
in \cite{Kla07} it is shown that weakly unbounded error quantum and classical CCs 
are within some constant factor (which is at most three \cite{INRYb07}). 
It turns out that the gap is a bit different in the QC model: 
there exists a Boolean function $f$ such that its classical weakly unbounded error QC 
is $\Omega(\log{n})$-times worse than its quantum correspondence. 
To show this, we will use a probabilistic method requiring the following Chernoff bound lemma 
from Appendix A of \cite{AS00}.

\begin{lemma}\label{chernoff}
Let $S=\{X_i\}$ be a set of $N$ independent random variables with $\Pr[X_i=1]=\Pr[X_i=-1]=\frac{1}{2}$. 
Then,
\[
\Pr\left[\left|\sum_{i=1}^N X_i\right| > a\right] < 2e^{-a^2/2N}.
\]
\end{lemma}

\begin{lemma}\label{wuqgap}\sloppy
There exists a partial Boolean function $f$ such that 
${WUC}(f) =\Omega(\log n)$ and ${WUQ}(f)=O(1)$.
\end{lemma}

\begin{proof}
We will again use the Fourier Sampling problem $\mathrm{FS}(F^a,g)=g_a$, which can clearly be solved exactly with a constant number of quantum queries for any choice of $g$. For the classical lower bound, we fix a string $g$ (to be determined shortly), and assume that $g$ is already known, so the algorithm need only make queries to $F^a$.

We will use the Yao principle \cite{Yao77} that the minimum number of queries required in the worst case for a randomized algorithm to compute some function $f$ with success probability at least $p$ for any input is equal to the maximum, over all distributions on the inputs, of the minimum number of queries required for a {\em deterministic} algorithm to compute $f$ correctly on a $p$ fraction of the inputs. Thus, in order to show a lower bound on the number of queries used by any randomized algorithm that succeeds with probability $1/2 + \beta$, it is sufficient to show a lower bound on the number of queries required for a deterministic algorithm to successfully output $g_a$ for a $1/2 + \beta$ fraction of the functions $F^a$ (under some distribution). We will use the uniform distribution over all inner product functions $F^a$ (recall that $F^a_x = \sum_i x_i \cdot a_i \mod 2$).

Now consider a fixed deterministic algorithm which makes an arbitrary sequence of $\frac{1}{3}\log n$ distinct queries to $F^a$, and then guesses the bit $g_a$. Assume without loss of generality that the algorithm makes exactly $\frac{1}{3}\log n$ queries on all inputs and that $n^{1/3}$ is an integer. There are at most $n^{1/3}$ possible answers to the queries, dividing the set $S$ of $n$ inner product functions into at most $k \le n^{1/3}$ non-empty subsets $\{S_i\}$, where $|S_i| \ge n^{2/3}$ for all $i$ (this is because each query will either split the set of remaining functions exactly in half, or will do nothing). Each subset will contain between 0 and $|S_i|$ functions $F^a$ such that $g_a=0$, with the remainder of the functions having $g_a=1$. For any $i$, define $m^i_0$ (resp. $m^i_1$) as the number of remaining functions $F^a \in S_i$ such that $g_a = 0$ (resp. $g_a = 1$). To succeed on the largest possible fraction of the inputs, the deterministic algorithm should guess the value with which the majority of the remaining bit strings in the subset $S_i$ picked out by the answers to the queries are associated. It is thus easy to see that this deterministic algorithm can succeed on at most the following fraction of the inputs.
\[ p = \frac{1}{2} + \frac{1}{2n} \sum_{i=1}^{k} |m^i_0-m^i_1|. \]
We now turn to finding a $g$ such that this expression is close to $1/2$ for all possible deterministic algorithms.

Our string $g$ will be picked uniformly at random from the set of all $n$-bit strings. This implies that, for an arbitrary {\em fixed} deterministic algorithm and for any $i$, $m^i_0$ and $m^i_1$ are random variables. Lemma \ref{chernoff} can thus be used to upper bound the fraction of the inputs on which this algorithm succeeds:
\[ \Pr[p > 1/2 + \beta] \le \Pr[\frac{1}{2n} \sum_{i=1}^{k} |m^i_0-m^i_1| > \beta] \le \Pr[\frac{1}{2 n^{2/3}} |m^1_0-m^1_1| > \beta] < 2e^{-2\beta^2 n^{2/3}}, \]
where it is sufficient for the bound to consider a fixed $i$ with $|S_i|=n^{2/3}$, w.l.o.g.\ assuming that this is true for $i=1$. The remainder of the proof is a simple counting argument. We find a rough upper bound on the number of deterministic algorithms using exactly $q$ queries on every input by noting that such an algorithm is a complete binary tree with $q + 1$ levels, where each leaf is labelled with 0 or 1 (corresponding to the output of the algorithm) and each internal node is labelled with a number from $[n]$ (corresponding to the input variable to query). There are thus fewer than $n^{2^{q+1}}$ deterministic algorithms using exactly $q$ queries. For $q=\frac{1}{3}\log n$, there are fewer than $2^{2n^{1/3}\log n}$ algorithms. We can now use a union bound to determine an upper bound on the probability $p'$, taken over random strings $g$, that {\em any} of these algorithms succeeds on a $1/2 + \beta$ fraction of the inputs.
\[ \Pr[p' > 1/2 + \beta] < 2^{2n^{1/3}\log n+1} e^{-2\beta^2 n^{2/3}} < 2 e^{2n^{1/3}(\log n-\beta^2 n^{1/3})}. \]
Let us pick $\beta = n^{-1/7}$. It can easily be verified that $\Pr[p' > 1/2 + \beta] < 1$ for sufficiently large $n$, so there exists {\em some} $g$ such that no classical algorithm that uses at most $\frac{1}{3} \log n$ queries can succeed on more than $1/2 + n^{-1/7}$ of the inputs. By Yao's principle, this implies that for this $g$, no randomized algorithm that uses at most $\frac{1}{3} \log n$ queries can solve $\mathrm{FS}(F^a,g)$ with a bias greater than $n^{-1/7}$.
Therefore, we have the desired separation: ${WUQ}(\mathrm{FS})=2$ (by Lemma \ref{eqfs}) while ${WUC}(\mathrm{FS}) = \Omega(\log n)$.
\end{proof}

This gap is asymptotically optimal, as we show with the following lemma.

\begin{lemma}\label{wuqlower}
For any Boolean function $f:\{0,1\}^n\rightarrow\{0,1\}$, 
${WUC}(f) \le 2 {WUQ}(f)\log{n}$.
\end{lemma}

\begin{proof}
Let ${\cal A}$ be an algorithm achieving ${WUQ}(f)$, 
i.e., ${\cal A}$ uses $d$ queries and has the success probability $1/2 + \beta$ 
such that ${WUQ}(f) = d + \log(1/2\beta)$. By the result of \cite{BBCMW98}, 
we know that there exists a polynomial that sign-represents $f$ such that 
its degree is $2d$, and its bias is $\beta$. Now we can use Lemma \ref{probcoeff}, 
which says that given such a polynomial, we can produce a randomized algorithm 
using at most $2d$ queries with success probability at least $1/2 + \beta/\sqrt{n^d}$. 
This implies that ${WUC}(f) \le 2d + \log(1/2\beta)  + d \log{n} \le 2 {WUQ}(f) \log{n} $.
\end{proof}

By Lemmas \ref{wuqgap} and \ref{wuqlower}, we obtain the following theorem.  

\begin{theorem}
There exists a partial Boolean function $f:\{0,1\}^n\rightarrow\{0,1\}$ 
such that $WUC(f)=\Theta(WUQ(f)\log n)$. 
\end{theorem}

\subsection{Gaps between $WUQ$ and $WUC$ for Total Functions}

In this subsection, we analyze the weakly unbounded error QC of some specific total Boolean functions. 
In contrast to the case for partial Boolean functions, our examples have only constant gaps between quantum and classical QCs.

\subsubsection{PARITY function}

It is straightforward to characterize the unbounded error QC of the function 
$\mbox{PARITY}:\{0,1\}^n \rightarrow \{0,1\}$, defined as $\mbox{PARITY}(x) = \oplus_i x_i$. 
It is a famous result of Minsky and Papert \cite{MP88} that ${sdeg}(\mbox{PARITY})=n$. 
With the algorithm of Lemma \ref{parity}, we therefore have ${UC}(\mbox{PARITY})={WUC}(\mbox{PARITY})=n$,
${UQ}(\mbox{PARITY})={WUQ}(\mbox{PARITY})=\lceil n/2 \rceil$.

\subsubsection{Threshold functions}\label{threshold}

First, let us consider the function $\mbox{OR}:\{0,1\}^n \rightarrow \{0,1\}$ 
which is defined as follows: $\mbox{OR}(x)=1$ iff $|x|\ge 1$, 
where $|x|$ is the Hamming weight of $x$. Consider the following 
single-query randomized algorithm for computing OR. 
Pick an input bit uniformly at random and query it. 
If the bit is $1$, output $1$. If the bit is $0$, output $1$ with probability $\frac{n-1}{2n-1}$, 
and $0$ otherwise. It is easy to see that this algorithm achieves bias $\frac{1}{4n-2}$, 
so we have ${UC}(\mbox{OR})=1$ and ${WUC}(\mbox{OR})\leq \log n + 2$.

In fact, by modifying the probability of outputting 1 properly, 
the above algorithm can also be used to compute threshold functions $\mbox{TH}_k$ 
defined by $\mbox{TH}_k(x) = 1$ if and only if $|x| > k$. 
Note that AND, OR, and MAJORITY are threshold functions. 
Without loss of generality, we can assume $k \le n/2 - 1$, 
since when $k \ge n/2$ one can consider the threshold function on flipped $x$. 
Now, the modified algorithm will output 1 with probability $q$, 
or otherwise, with probability $1-q$ query $x$ at random position, 
say, $i$, and output the value of $x_i$. 
Therefore, choosing $q = (1/2 - r)/(1-r)$ for $r = (k+1/2)/n$, if $|x| \le k$, 
then the probability of outputting 1 is at most $q + (1-q)\frac{k}{n} < 1/2$. 
Otherwise, it is at least $q + (1-q)\frac{k+1}{n} > 1/2$. 
Thus, we have an unbounded error algorithm for $\mbox{TH}_k$. 
Moreover, it is easy to see that the bias is $\Omega(1/n)$, 
and therefore to conclude that ${WUC}(\mbox{TH}_k)\ = O(\log n)$.

On the other hand, we can lower bound $WUQ(f)$ for {\em any} non-constant symmetric function $f$ 
using the polynomial method. Let $p$ be a degree $d$ unbounded error polynomial representing 
$f$ with bias $\beta$ and $0 \le p(x) \le 1$ for all $x \in \{0,1\}^n$. 
By Lemma \ref{degquery}, ${WUQ}(f)$ can be bounded 
in terms of a tradeoff between $d$ and $\beta$, using techniques of \cite{NS94}, 
which are based on the following well-known lemma of Ehlich and Zeller \cite{EZ64} 
and Rivlin and Cheney \cite{RC66}:
\begin{lemma}\label{degbound}
Let $p$ be a degree $d$ polynomial such that, 
for any integer $0\le i\le n$, $b_1 \le p(i) \le b_2$, and for some real $0 \le x \le n$, 
$|p'(x)| \ge c$. Then $d \ge \sqrt{cn/(c+b_2-b_1)}$.
\end{lemma}
In order to use this lemma, we first note that $p$ can be symmetrized \cite{MP88,NS94} 
to produce a univariate polynomial $q$ of degree at most $d$ defined  
via the following mapping: $q(x) = \left(\sum_{y,|y|=x} p(y) \right)/\binom{n}{x}$. 
Since $f$ is not constant, there exists a $k\in\{0,1,\ldots,n\}$ such that 
$q(k)\le 1/2-\beta$, and that either $q(k-1)\ge 1/2+\beta$ or $q(k+1)\ge 1/2+\beta$. 
Thus, there must exist some $x$ in $[k-1,k]$ (or in $[k,k+1]$) such that 
$|q'(x)|\ge 2 \beta$. By Lemma \ref{degbound}, $d \ge \sqrt{2 \beta n/(2 \beta + 1)}$, 
which implies that
\[ {WUQ}(f) \ge \min_\beta \left(\sqrt{\frac{n \beta}{4\beta + 2}} + \log(1/\beta)\right). \]
To simplify this expression, we note that the elementary inequality $\sqrt{4+2/\beta} 
\le 1+1/\beta$ (for $0 < \beta < 1/\sqrt{3}$) gives
\[ {WUQ}(f) \ge \min_\beta \left(\frac{\sqrt{n}}{1+1/\beta} + \log(1/\beta)\right). \]
By minimizing this expression over $\beta$ we see that the minimum is found at
\[ \beta = \frac{1}{2} \left( \sqrt{n} \ln{2}  - 2 - \sqrt{n (\ln 2)^2  - 4 \sqrt{n}\ln 2 } \right). \]
Now we can use the series expansion of the square root function to upper bound $\beta$ as follows:
\begin{eqnarray*}
\beta 
&=& \frac{1}{2} \left( \sqrt{n} \ln 2  - 2 - \sqrt{n} \ln 2  \sqrt{1-4/(\sqrt{n}\ln 2) } \right) \\
&=& \frac{1}{2} \left( \sqrt{n} \ln 2  - 2 - \sqrt{n} \ln 2 \left(1 - 2/(\sqrt{n})\ln 2  - 2/(\ln 2)^2 n + O(n^{-3/2}) \right) \right) \\
&<& \frac{1}{\sqrt{n}\ln 2 }.
\end{eqnarray*}
Given this upper bound on $\beta$, it is immediate that ${WUQ}(f) \ge \log(1/2\beta) \ge (\log n)/2 - O(1)$. 

Now we summarize the results on the unbounded error and weakly unbounded error QCs 
of the threshold function. 
\begin{theorem}\label{thm:threshold}
${UC}(\mathrm{TH}_k)={UQ}(\mathrm{TH}_k)=1$ and ${WUC}(\mathrm{TH}_k) = {WUQ}(\mathrm{TH}_k) = \Theta(\log n)$.
\end{theorem}

\subsection{Other Complexity Measures}

Can we relate $UQ$ or $WUQ$ to any other interesting complexity measures of Boolean functions \cite{BW02}? 
One might hope to show that some well-studied property of Boolean functions gives a lower bound on $UQ$. 
One of the weakest such measures is {\em average sensitivity} (also known as {\em total influence}). 
The sensitivity of a function $f:\{0,1\}^n\rightarrow\{0,1\}$ 
at input $x$ is defined as $s_x(f) = \sum_{i\in[n]} |f(x) - f(x \oplus e_i)|$, 
where $e_i$ is the bit string with 1 at position $i$, and 0 elsewhere. 
The average sensitivity of $f$ is the average over all $x$: $\overline{s}(f) = \left( \sum_x s_x(f) \right) / 2^n$.

$\overline{s}(f)$ is a lower bound on many other interesting complexity measures, such as block sensitivity and certificate complexity \cite{BW02}. In particular, Shi \cite{Shi00} has shown that $\overline{s}(f)$ is a lower bound on the bounded error quantum QC of $f$. 
However, we now show that $\overline{s}(f)$ can be exponentially larger than even ${WUC}(f)$. 
This implies that the unbounded error complexity models studied in this paper are somehow 
too weak to be comparable with the usual complexity measures of Boolean functions. 
The example we use is simply the threshold function $\mbox{TH}_{n/2}$, or in other words the
MAJORITY function.

\begin{lemma}\label{as}
Assume $n$ is even. Then ${WUC}(\mbox{TH}_{n/2})=O(\log n)$ 
while $\overline{s}(\mbox{TH}_{n/2})=\Omega(\sqrt{n})$.
\end{lemma}

\begin{proof}
The first half follows from the discussion at the start of Section \ref{threshold}. The second half is folklore; for an explicit proof, note that $s_x(\mbox{TH}_{n/2})=0$ unless $|x|=n/2$ or $|x|=n/2+1$. When $|x|=n/2$, $s_x(\mbox{TH}_{n/2})=n/2$, and when $|x|=n/2+1$, $s_x(\mbox{TH}_{n/2})=n/2+1$. Thus
\begin{eqnarray*}
\overline{s}(\mbox{TH}_{n/2}) &=& \dfrac{1}{2^n} \left( \binom{n}{n/2} \dfrac{n}{2} + \binom{n}{n/2+1} \left( \dfrac{n}{2} + 1 \right) \right) \ge \dfrac{n}{2^{n+1}} \binom{n}{n/2} \ge \dfrac{\sqrt{n}}{2 \sqrt{\pi}}
\end{eqnarray*}
where we use Stirling's approximation.
\end{proof}

\section{Concluding Remarks}
We have completely characterized the unbounded error quantum QC as half of its classical counterpart, 
and have given a lower bound on the weakly unbounded error quantum QC which is tight for partial functions. 
However, some open questions remain. 
For example, for total functions $f$, is it the case that ${WUC}(f)=O({WUQ}(f))$? 
One might expect this to be true as total functions do not have big gaps between quantum and classical QCs 
in the bounded error setting: there can be at most a polynomial separation between the quantum and classical 
QCs of total functions \cite{BBCMW98} while a partial function gives us an exponential gap between them. 
It is also intriguing to note that the factor of 2 separation between $UQ$ and $UC$ 
is the same as the maximal known separation between the exact quantum and classical QCs 
of total Boolean functions -- perhaps the techniques here could provide insight into whether this is optimal.

\section{Acknowledgements}

AM was supported by the EC-FP6-STREP network QICS, and would like to thank Richard Low for helpful discussions, and in particular for help in simplifying the proof of Theorem \ref{thm:threshold}. HN was supported in part by Scientific Research Grant, Ministry of Japan, 19700011.

\end{document}